\documentclass[useAMS,usenatbib,usegraphicx]{mn2e}
\usepackage{amssymb}
\usepackage{float}
\usepackage[font=footnotesize,caption=false]{subfig}
\usepackage{booktabs}
\usepackage[usenames,dvipsnames]{color}

\newcommand{\mnras}{MNRAS}
\newcommand{\nat}{Nat}
\newcommand{\aj}{AJ}
\newcommand{\apj}{ApJ}
\newcommand{\apjl}{ApJL}
\newcommand{\apjs}{ApJS}

\newcommand{\aap}{A\&A}

\newcommand{\pasp}{PASP}
\newcommand{\araa}{ARA\&A}
\newcommand{\physrep}{Phys. Rep.}

\def\Msun{\hbox{M$_{\odot}$}}
\def\sfrunits{\Msun\ yr$^{-1}$}
\def\halpha{H$\alpha$  }
\def\Lha{$L_{\mathrm{H}\alpha}$}

\def\NII{[N\hspace{.03cm}II]}

\def\OII{[O\hspace{.03cm}II]}
\def\OIII{[O\hspace{.03cm}III]}
\def\HII{H\hspace{.03cm}{\small II}~}
\def\LIR{$L_\mathrm{IR}$}
\def\Td{$T_\mathrm{d}$}
\def\Lsun{\hbox{L$_{\odot}$}}
\def\mum{$\mu$m}

\title[Dust Properties of Clumpy Disc Galaxies at z$\sim$1.3]{Dust Properties of Clumpy Disc Galaxies at z$\sim$1.3 with Herschel-SPIRE\thanks{{\it Herschel} is an ESA space observatory with science instruments provided by European-led Principal Investigator consortia and with important participation from NASA.}}
\author[Wisnioski et al.]{Emily Wisnioski$^{1,2}$\thanks{E-mail: emily@mpe.mpg.de},
Karl Glazebrook$^{2}$, 
Chris Blake$^{2}$,
Mark Swinbank$^{3}$\\
$^{1}$Max-Planck-Institut f\"{u}r extraterrestrische Physik, Postfach 1312, Giessenbachstr., D-85741 Garching, Germany.\\
$^{2}$Centre for Astrophysics and Supercomputing, Swinburne University of Technology, P.O. Box 218, Hawthorn, VIC 3122, Australia.\\
$^{3}$Institute for Computational Cosmology, Durham University, South Road, Durham DH1 3LE, UK.}

\date{Accepted 2013 August 17. Received 2013 August 16; in original form 2013 July 10}

\pagerange{\pageref{firstpage}--\pageref{lastpage}} \pubyear{2013}

\begin{document}
\label{firstpage}
\maketitle

\begin{abstract}
We present the far-infrared derived dust properties from \textit{Herschel} SPIRE of the WiggleZ kinematic sample of 13 star-forming galaxies at $z\sim1.3$, with existing ancillary $\sim$kpc resolution integral field spectroscopy. We detect 3 galaxies individually and place limits on the remainder by stacking. The detected galaxies, two clumpy discs and one merger, have cold dust temperatures of \Td~$\sim 26$ K and have infrared luminosities of \LIR~$\sim 1.2\times10^{12}$ \Lsun, determined by modified blackbody fitting. The two detected disc galaxies have the largest \halpha surface areas of the sample and have the reddest ultraviolet to near-infrared spectral energy distributions. The likely source of the infrared luminosity in these objects is dust heated by the interstellar radiation field and young stellar emission from the clumps within the discs. The source of infrared luminosity for the merger is likely a dust heated by a starburst resulting from the merger. The WiggleZ detections are among the coldest and lowest luminosity individual objects detected in the far-infrared at $z>1$. When combining the kinematic data, we find that none of the compact galaxies nor the  `dispersion dominated' galaxies of the WiggleZ kinematic sample are detected, implying that they have warmer dust temperatures. The compact objects show the highest H$\alpha$ velocity dispersions in the sample, in qualitative agreement with bulge formation models. These far-infrared results strengthen the interpretation that the majority of galaxies in this sample constitute different stages in clumpy disc formation as presented from ancillary kinematic analysis. \\

\end{abstract}

\section{Introduction}
The majority of results from $z\sim1$ galaxy formation studies have utilised observations of the rest-frame optical where key physical and chemical parameters can be measured from nebular emission lines, e.g. star formation rates (SFRs) and metallicity (from \halpha, H$\beta$, \NII, \OII, \OIII). However, these lines can be strongly attenuated by intervening dust within galaxies. Locally, where high spatial resolution studies of dusty galaxies has been possible, prescriptions to account for the global effects of dust have been calibrated.  For studies of star-forming galaxies at $z>1$ estimates of dust extinction are either determined using calibrations set from local analogs or spectral fits to broad-band photometry. However, both methods are highly uncertain as in the first case there the local relations may not hold for the early universe \citep{2009ApJ...703.1672K} and in the later case estimates can be severely limited by degeneracies between dust extinction and age (e.g. \citealt{1997ApJ...487..625G}). 
Accurate dust corrections at high-redshift will be especially important for upcoming galaxy surveys from new near-infrared instruments that will rely on \halpha for SFR estimates.  Furthermore, it has been suggested that star-forming galaxies (SFGs) were dustier in the past (e.g. \citealt{2005A&A...440L..17T}) which if not well understood could lead to misinterpretation of the true evolution in galaxy properties.

The launch of modern far-infrared (FIR) facilities, including the \textit{Herschel Space Observatory} \citep{2010A&A...518L...1P}, has made it possible to directly investigate dust properties of star-forming galaxies from $z\sim0-3$.  With past facilities only the brightest objects were detectable, making it difficult to study dust in a representative sample of star-forming galaxies at high redshift. The galaxies studied were typically classified as mergers, starburst galaxies or submillimeter galaxies (e.g. \citealt{2008ARA&A..46..201S}) that lie above the main locus of galaxies in the SFR-stellar mass plane \citep{2007ApJ...670..156D}.
\textit{Herschel} has allowed the first direct detections of more representative star-forming, or `main sequence', galaxies at $z\gtrsim1$ (e.g. \citealt{2011A&A...533A.119E,2011ApJ...734L..12B,2012MNRAS.426.1782R, magnelli:2012aa,2013arXiv1306.1121O}). Measurements of dust extinction in high redshift galaxies are made from these observations by studying the relationship between infrared SFRs and \halpha SFRs. A comparison of the ratio of these two SFR indicators with additional methods of measuring dust extinction yield a good agreement between methods using $\sim$500 galaxies at $z=0.06-0.46$, covering a wide range of masses and SFRs \citep{dominguez:2012:05}, and $\sim$60 far-infrared selected sources at $z\sim1$ of star-forming galaxies \citep{2012MNRAS.426.1782R}, suggesting that local calibrations are appropriate to use for high redshift galaxies.  However, at both redshifts a correlation is seen between metallicity and the ratio of infrared to \halpha SFRs that may be redshift dependent. 

Results from studies locally and at high redshift, where ancillary data is available, indicate a correlation between infrared luminosities and morphology and/or galaxy type (e.g. \citealt{1988ApJ...325...74S,1996ARA&A..34..749S,2004PhDT........18I,2010ApJ...721...98K,2011arXiv1110.4057K}). 
Statistical results from $z=0.1-3.5$ suggest that \LIR~$\sim5\times10^{11}$ \Lsun~is the quantitative dividing luminosity between the dominance of spirals and major mergers, comparable to current detection limits of \textit{Herschel} at $z\sim1$ \citep{2011arXiv1110.4057K}. 
Thus, despite the higher sensitivity of the new space-based facilities compared to ground-based facilities, observations of large samples of star-forming galaxies (SFGs) at $z\gtrsim1$ have remained a challenge. Long integrations are needed, severely limiting large surveys of this population and confining searches for high-redshift detections to the \textit{Hubble} deep fields. Stacking is still often required to derive infrared properties of UV-selected galaxies at $z>1$ \citep{2010ApJ...720L.185M,2012ApJ...744..154R,2013arXiv1307.3556I}. 

Dust temperatures derived from far infrared photometry have also been shown to correspond on average with morphology and galaxy type \citep{2011A&A...533A.119E,magnelli:2012aa}.  Star-forming galaxies at $0<z<2.5$ have cooler effective dust temperatures ($\sim31$ K) than starburst galaxies or submillimeter galaxies (SMGs; $\sim40$ K; \citealt{2011A&A...533A.119E,magnelli:2012aa}). Galaxies with hot dust temperatures tend to be more compact \citep{2011A&A...533A.119E} and have disturbed morphologies indicative of major mergers, while galaxies with cold dust temperatures have disc-like morphologies \citep{magnelli:2012aa}.  However, even with deep space-based imaging, at $z>1$ morphologies are hard to classify kinematically due to the high number of irregular galaxies which confuse the simple disc vs. merger classification. 

This paper presents far-infrared observations of 13 galaxies with ancillary \halpha kinematics from integral field spectroscopy in order to directly compare dust properties with global galaxy motions.  The observations and derivation of fluxes from \textit{Herschel}/SPIRE for the WiggleZ kinematic sample are described in Section~\ref{sec5.obs}. Modified blackbody fits are applied in Section~\ref{sec5.results} to estimate total infrared luminosities, dust temperatures, and dust masses. The dust extinction derived from the ratio of \halpha and infrared luminosities is compared to that derived from SED fitting to ultraviolet (UV) to near-infrared (NIR) photometry. Section~\ref{sec5.discussion} looks at the correlation between kinematic properties and dust properties, re-evaluating the galaxy classifications made in \cite{2011MNRAS.417.2601W}. A standard $\Lambda$CDM cosmology of $\Omega_{\mathrm{m}} = 0.27$, $\Omega_{\Lambda} = 0.73$, $h = 0.71$ is adopted throughout this paper with a Baldry-Glazebrook initial mass function (IMF; BG03; \citealt{2003ApJ...593..258B}). In this cosmology, at redshift $z=1.3$, 1 arcsec corresponds to 8.44 kpc.

\vspace{-0.2cm}\section{Observations and data}
\label{sec5.obs}
We present FIR observations for 13 UV-selected galaxies at $z\sim1.3$ selected from the \textit{WiggleZ Dark Energy Survey} \citep{Drinkwater:2010bx} and observed using integral field spectroscopy (IFS) with the OSIRIS instrument on the Keck telescopes \citep{2011MNRAS.417.2601W}. 
The galaxies in this sample are ideal targets for \textit{Herschel} because they are among the most luminous UV-selected sources at $z\sim1.3$. They were originally selected as being possible starbursts due to ULIRG-like star formation rates measured from their optical spectra \citep{2011MNRAS.417.2601W}. A fraction of the sample (7/13) have subsequently been found to be more disc-like, representing the luminous end of main-sequence star-forming galaxies with SFR$_{\mathrm{H}\alpha}$ $\sim40$ \sfrunits. Thus, although the sample is UV-selected, suggesting low dust extinction, the high \halpha luminosities (more typical of $z\sim2$ galaxies) indicate that these objects are detectable by \textit{Herschel} if $A_V\sim1$, normally assumed for emission line studies of SFGs (e.g. \citealt{1992ApJ...388..310K,1996MNRAS.281..847T}) and inline with estimates from their spectral energy distributions (SEDs). Furthermore, 5/13 of the galaxies appear to be compact similar to SMGs with chaotic kinematics possibly representing the strong starbursts resulting from major mergers (e.g. \citealt{2008MNRAS.391..420S,2010ApJ...724..233E}). Finally, at $z\sim1.3$ the three \textit{Herschel} bands at 250, 350, and 500 $\mu$m would straddle the peak of the infrared SED allowing constraints on the dust temperature.

\begin{table*}
\begin{minipage}{\textwidth}
\caption{Summary of WiggleZ galaxy properties}
\begin{tabular*}{\textwidth}{@{\extracolsep{\fill}}llcccccccc}
\hline
{WiggleZ ID	} & 
{ID	} & 
{z } & 
{SFR$_{\mathrm{H}\alpha}^a$} & 
{$\sigma^a$} & 
{$\log$(M$_{*}$[\Msun])$^b$ } &  
{$A_V^b$} &
{Morphological} \\
{} & 
{} & 
{} & 
{(\sfrunits)} & 
{(km s$^{-1}$)} & 
{} &
{(mag)} &
{Type$^c$}  \\
\hline
R03J032450240$-$13550943 & WK0912\_13R &  1.2873 &  20.3 $\pm$ 6.8 & ~81.6 $\pm$ 27.7  &  10.7  $\pm$  0.2 & 1.1 $\pm$ 0.3 &~M$^d$\\
S15J145355248$-$00320351 & WK1002\_61S &  1.3039 &  35.6 $\pm$ 2.3 & ~85.1 $\pm$ 20.3  &  10.7  $\pm$  0.7 & 0.2 $\pm$ 0.2 &M\\
R01J005822757$-$03034040 & WK0909\_02R &  1.3616 &  20.2 $\pm$ 3.8 & ~87.9 $\pm$ 21.9  &  10.3  $\pm$  0.6 & 0.7 $\pm$ 0.4 &M\\
R03J032206214$-$15443471 & WK0909\_06R &  1.4602 &  54.8 $\pm$ 3.9 & ~92.6 $\pm$ 27.8  &  11.0  $\pm$  0.2 & 0.8 $\pm$ 0.3 &~M$^d$\\
                                                                                                     
S15J144102444+05480354   & WK0905\_22S &  1.2810 &  44.0 $\pm$ 1.8 & ~98.2 $\pm$ 10.9  &  10.3  $\pm$  0.2 & 0.3 $\pm$ 0.2 &~E$^d$\\
R00J232217805$-$05473356 & WK0912\_01R &  1.2960 &  24.6 $\pm$ 3.2 & ~98.6 $\pm$ 19.6  &  10.0  $\pm$  0.6 & 0.5 $\pm$ 0.2 &~E$^d$\\
S09J091517481+00033557   & WK1002\_41S &  1.3330 &  29.9 $\pm$ 2.0 & ~91.4 $\pm$ 13.0  &  10.2  $\pm$  0.1 & 0.4 $\pm$ 0.1 &~E$^d$\\
S00J233338383$-$01040629 & WK0809\_02S &  1.4521 &  42.5 $\pm$ 7.7 & ~92.7 $\pm$ 16.6  &  11.1  $\pm$  0.6 & 0.0 $\pm$ 0.1 &E\\
                                                                                                     
S15J142538641+00483135   & WK1002\_18S &  1.3044 &  25.3 $\pm$ 2.1 & 102.1 $\pm$ 20.0 &  10.2  $\pm$  0.1 & 0.5 $\pm$ 0.1 &~S$^d$\\
S11J101757445$-$00244002 & WK1002\_14S &  1.3048 &  40.6 $\pm$ 1.9 & 113.7 $\pm$ 8.8~  &  10.6  $\pm$  0.2 & 0.3 $\pm$ 0.1 &~S$^d$\\
S09J090933680+01074587   & WK1002\_46S &  1.3074 &  22.0 $\pm$ 2.0 & 124.8 $\pm$ 24.9 &   ~9.8  $\pm$  0.6 & 0.2 $\pm$ 0.3 &S\\
S11J110405504+01185565   & WK1002\_52S &  1.3191 &  38.6 $\pm$ 2.7 & 125.4 $\pm$ 20.9 &  11.7  $\pm$  0.1 & 0.0 $\pm$ 0.1 &S\\
S09J090312056$-$00273273 & WK0912\_16S &  1.3260 &  17.6 $\pm$ 2.0 & ~85.1 $\pm$ 18.6  &   ~9.8  $\pm$  0.2 & 0.1 $\pm$ 0.1 &S\\
\hline
\label{table.global5}
\end{tabular*}\\
$^a$ \halpha and kinematic properties are derived from OSIRIS near-infrared observations with a BG03 IMF.\\
$^b$ Values derived from SED fitting to UV, optical, and NIR magnitudes with a BG03 IMF.\\
$^c$ Morphological classification based on 0.05$''$ \halpha imaging with OSIRIS; M: multiple emission, E: extended emission, S: single emission.
$^d$ \halpha velocity map consistent with a model exponential disc.
\end{minipage}
\end{table*}

\vspace{-0.2cm}\subsection{OSIRIS sample}
\label{sec5.rehash}
The \halpha morphological classifications defined in \cite{2011MNRAS.417.2601W} are discussed here in the context of the \textit{Herschel} observations. In brief, the sample can be broken down into three morphological classes based on their \halpha images; `single emission galaxies' appear as resolved compact galaxies, `extended emission galaxies' typically have a bright knot of emission with tails or accompanying diffuse light, and `multiple emission galaxies' show multiple resolved knots of \halpha emission connected by diffuse light. These classifications and a summary of key galaxy properties derived from optical and near infrared (NIR) data are reproduced in Table~\ref{table.global5}.

Exponential disc models were fit to the observed \halpha velocity maps of the full sample utilising a reduced chi-squared minimisation with Monte Carlo Markov Chain methods. Seven galaxies were found to be consistent with an exponential disc model when taking into account the chi-squared value of the fit and the residual of the difference between the model velocity map and observed velocity map. These fits and associated uncertainties are discussed in more detail in Section 3.6 of \cite{2011MNRAS.417.2601W}. The galaxies consistent with a disc model are indicated in Table~\ref{table.global5}. One galaxy, WK0905\_02R, was identified as a possible merger in the Appendix of \cite{2011MNRAS.417.2601W} due to a low surface-brightness optical counterpart identified at the same redshift, which extends out to 21 kpc from the brightest region of WK0905\_02R detected with OSIRIS.

Dust extinctions, $A_V$, were estimated for all the objects in the sample from the best-fit SED using UV, optical, and NIR photometry assuming a Calzetti dust law \citep{2000ApJ...533..682C} and are given in Table~\ref{table.global5}. This is a common technique for high-redshift galaxies however, these values can be uncertain due to poor sampling of the full SED, as will be investigated in Section~\ref{sec5.dust}. 

\subsection{Herschel observations}
\label{subsec5.herschel}
From this sample,  all 13 galaxies have been observed for 1687 s with the \textit{Herschel Spectral and Photometric Imaging REceiver} (SPIRE; \citealt{2010A&A...518L...3G}) imager in the 250, 350 and 500 $\mu$m bands.  The pixel size and beam FWHM for SPIRE are 6$''$, 10$''$, 14$''$ and 18.2$''$, 24.9$''$, 36.3$''$ respectively for the 250, 350 and 500 $\mu$m bands. The images are calibrated using the standard SPIRE pipeline in units of mJy/beam. The confusion limit for each band is 5.8, 6.3, and 6.8 mJy/beam. The total variance in each image is estimated by adding in quadrature the confusion noise, $\sigma_\mathrm{conf}$, and the instrument noise over the observation time as given by \cite{2010A&A...518L...5N}. 

The reduced images are calibrated such that the source flux is given by the flux in the pixel at the peak of the PSF at the expected location of the source \citep{2010A&A...518L...5N}. The fluxes of the detected WiggleZ sources were extracted by fitting a 2D Gaussian, taking the peak as the total flux as outlined in the \textit{Source Fitting of Point Sources in Maps} recipe of the SPIRE Photometry Cookbook\footnote[1]{The SPIRE Photometry Cookbook (Bendo et al. 2011; version July 2011) available at: http://herschel.esac.esa.int/hcss-doc-10.0\\ /load/spire\_drg/html/spire-photometer.html} and are given in Table~\ref{table.herschel}. The extraction includes division by the pixelisation correction factor defined in the SPIRE Observers' Manual\footnote[2]{The SPIRE Observers' Manual available at:\\ http://herschel.esac.esa.int/Docs/SPIRE/pdf/spire\_om.pdf} which corresponds to $P=[0.951,0.931,0.902]$ respectively for $[250,350,500]$ $\mu$m. Additionally, the appropriate correction has been made to the 350 $\mu$m fluxes in the cases in which the data was reduced with the out of date version of the reduction pipeline HIPE V5$^1$. Errors on source flux values are given by adding in quadrature the total image variance with the error on the peak flux value determined from the 2D Gaussian fitting with the error extension used as the corresponding error image in the fit.

\begin{figure*}
\includegraphics[scale=0.9,  trim=0cm 18.5cm 0cm 0cm, clip ]{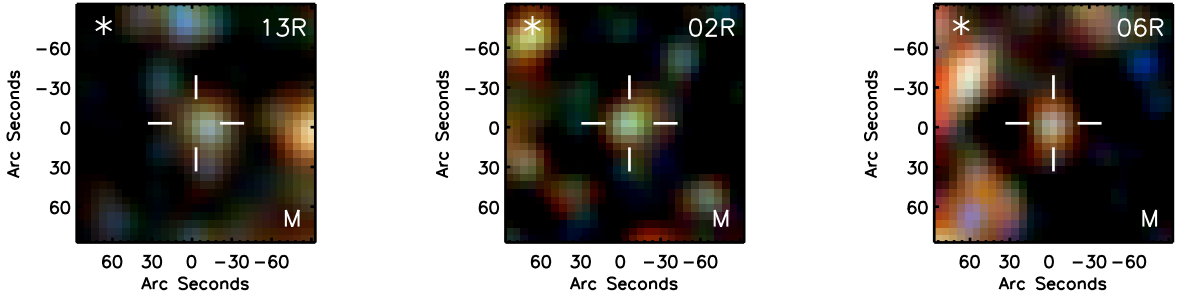}
\caption[RGB images created from SPIRE bands ($250\mu$m, $350\mu$m, $500\mu$m)]{RGB images made from the three SPIRE bands ($250\mu$m : $blue$, $350\mu$m : $green$, $500\mu$m : $red$). Each image is 3 arc min on a side. The images are centred on the expected position of the WiggleZ galaxies based on optical astrometry and confirmed by high resolution near-infrared images. The \halpha morphology is given in Table~\ref{table.global5} and denoted in the bottom right corner of each image. }
\label{fig.rgbSPIRE}
\end{figure*}

Detections at $3\sigma$ confidence are defined by the flux limit $3\times\sigma_\mathrm{conf}$, or $>17.4$ mJy, $>18.9$ mJy, and $>20.4$ mJy for the 250, 350 and 500 $\mu$m bands. Only three of the thirteen objects observed, WK0909\_02R, WK0909\_06R, and WK0912\_13R, are detected in more than two wavebands. For non-detected sources upper-limits are quoted as $3\sigma_\mathrm{conf}$. Tri-colour images created from the three SPIRE bands are shown in Fig.~\ref{fig.rgbSPIRE} for the detections. Fluxes for the detected galaxies are given in Table~\ref{table.herschel}.

\begin{table*}
\begin{minipage}{\textwidth}
\caption{\textit{Herschel} SPIRE observations and dust properties of WiggleZ galaxies}
\begin{tabular*}{\textwidth}{@{\extracolsep{\fill}}lcccccccc}
\hline
{ID	} & 
{f$_{250}$} &
{f$_{350}$} &
{f$_{500}$ } &
{log(L$_\mathrm{IR}$[L$_{\odot}$])  } & 
{T$_\mathrm{dust}$ } & 
{log(M$_\mathrm{dust}$[\Msun]) } & 
{A$_{\mathrm{H}\alpha}^b$}&
{SFR$_\mathrm{corr}^{\mathrm{H}\alpha,b}$}\\
{} &
{(mJy)} & 
{(mJy)} &
{(mJy)}  &
{} & 
{(K)} &
{} &
{(mag)} &
{(\sfrunits)}\\
\hline
WK0912\_13R$^a$    & $18.72 \pm 5.93$  & $25.53   \pm 6.43$  &  $17.63 \pm 6.99$   &  $11.8\pm0.1$  & $21.6 \pm 4.4$    & $9.4\pm0.4$ & 0.92 & 48 \\
WK0909\_02R           & $25.85 \pm 5.90$  & $21.46  \pm  6.40$  &  $12.02 \pm  6.95$  &  $12.1\pm0.3$   & $29.8 \pm 8.4$   & $9.0\pm1.3$ & 1.40  & 73\\
WK0909\_06R           & $36.09 \pm 5.90$  & $33.47  \pm  6.39$  &  $23.22 \pm  6.92$  &  $12.3\pm0.1$   & $27.5 \pm 4.0$   & $9.3\pm0.2$ &  0.96 & 132\\
\hline
\label{table.herschel}
\end{tabular*}\\
$^a$ Objects with FIR properties derived from deblended sources as outlined in Section~\ref{subsec5.wise}. \\
$^b$ Derived from the ratio of infrared and \halpha luminosities using the prescription of \citet{2009ApJ...703.1672K}. \\
\end{minipage}
\end{table*}

\subsection{Deblending with WISE}
\label{subsec5.wise}
\begin{figure}
\begin{center}\includegraphics[scale=0.58,  trim=5.5cm 7.5cm 0cm 8cm, clip]{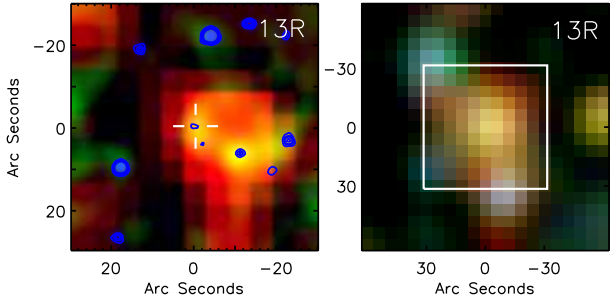}
\end{center}
\vspace*{-0.2cm}\caption[Deblended RGB image of WK0912\_13R created from SPIRE bands]{\textit{(left)} RGB image made from SPIRE $250\mu$m ($red$), WISE W$2-4.6\mu$m ($green$), and near-infrared IRIS2 Ks ($blue$ contours) images of WK0912\_13R. A bright nearby source confuses the possible detection of the WiggleZ target as seen by the bright 4.6 \mum~source $\sim15''$ south-west of the expected location of the WiggleZ galaxy. \textit{(right)} Deblended RGB image of WK0912\_13R created from SPIRE bands ($250\mu$m, $350\mu$m, $500\mu$m). The neighbouring source was modelled and removed in each SPIRE band to create this deblended image of the galaxy. The remaining flux is detected at the WiggleZ position above $3\sigma_\mathrm{conf}$ in the 250 and 350 \mum~bands. The white box indicates the zoomed-in region shown in the \textit{left} image}
\label{fig.rgb13R}
\end{figure}

Due to the large pixels and beam sizes, many astronomical sources may overlap within one SPIRE beam. The near neighbours can contaminate source extraction giving inaccurate flux measurements. Higher resolution 4.6 $\mu$m images, with spatial resolution of $6.4''$/pixel, from the \textit{Wide-field Infrared Survey Explorer} (WISE; \citealt{2010AJ....140.1868W}) all-sky survey \citep{2012yCat.2311....0C} are used as a prior to determine precise locations of sources nearby to the target WiggleZ galaxies (e.g. \citealt{2006ApJ...649L..67L,2007ApJ...655...51W,2011ApJ...735...86W}), and to check the \textit{Herschel} and OSIRIS astrometry. Within the all-sky WISE data release the three \textit{Herschel} detected galaxies from this sample are detected in at least one of the four WISE wavebands. Of the galaxies detected with \textit{Herschel} only one, WK0912\_13R, is confused by nearby sources in the \textit{Herschel} images. Figure~\ref{fig.rgb13R} shows the tri-colour image of WK0912\_13R using near infrared IRIS2 Ks-band images (\textit{blue contours}), WISE 4.6 \mum~images (\textit{green}), and SPIRE 250 \mum~images (\textit{red}). The 4.6 \mum~and 250 \mum~images are selected for this image because they have the highest resolution and highest detection fraction for the WiggleZ kinematic sample. 

An attempt is made to deblend the nearby sources from the expected location of WK0912\_13R using the WISE 4.6\mum~image. In the left panel of Fig~\ref{fig.rgb13R}, two WISE sources are co-located with the SPIRE 250 \mum~beam near the expected location of the WiggleZ galaxy.  Assuming these two sources contribute to the 250 \mum~emission, one being the WiggleZ source and the other a bright nearby source $\sim15''$ south-west of the expected location of the WiggleZ galaxy. These locations are then fixed within the image and a SPIRE source is modelled at each location by a two-dimensional Gaussian with FWHM of the SPIRE beam to represent the PSF of these images. A Gaussian provides a good estimate of the shape of the SPIRE beam. A chi-squared minimisation of the difference of the data and the model is conducted using downhill simplex methods with the flux of each modelled source as the only free parameters.  The model image of the neighbouring source is then subtracted from the actual image and a flux is extracted from the residual at the expected WiggleZ position by the same method described in Section~\ref{subsec5.herschel}. The deblended image of WK0912\_13R is shown in the right panel of Fig.~\ref{fig.rgb13R}. The tricolour SPRIE image reveals a similar colour as WK0909\_02R and WK0909\_06R, suggesting that the deblending has been successful. The resulting fluxes are $3.2\sigma_\mathrm{conf}$, $4.1\sigma_\mathrm{conf}$, and $2.6\sigma_\mathrm{conf}$ in the 250, 350, and 500 \mum~bands respectively, and are given in Table~\ref{table.herschel}.

\subsection{Stacking}
\label{subsec5.stacking}
The ten non-detected sources were stacked to put a limit on the typical infrared luminosity and dust content on the remainder of the WiggleZ kinematic sample. 
The stacking technique follows from \cite{2010A&A...518L..25R} and \cite{2010ApJ...720L.185M} and is described shortly here. First, $60''\times60''$ images were extracted around each object in the three SPIRE bands. 
Because of the large pixel sizes of the SPIRE images, sub-pixel shifts were applied in an effort to best recover the stacked emission without smearing a detection. The shifts were performed using a Fourier transform interpolation. To avoid image defects and correlations arising from the coverage pattern each image was rotated $90^\circ$ relative to the previous image in the stack. The images were then median-combined to construct a final stacked image in each band. Signal is detected in the 250 \mum~band stack at 6.1 mJy. Perhaps surprisingly, no flux is detected above the confusion noise in the stacks of the 350 or 500 \mum~bands. This is attributed to the low number statistics of the sample. Upper limits are derived in each band at $z\sim1.32$ from the stacked images resulting in $<5.2$ mJy and $<5.1$ mJy for the 350, and 500 $\mu$m bands respectively.

\section{Results \& Analysis}
\label{sec5.results}

\subsection{Detections vs. non-detections}
\label{subsec5.detections}
\begin{figure}
\begin{center}\hspace*{-0.5cm}\includegraphics[scale=0.5]{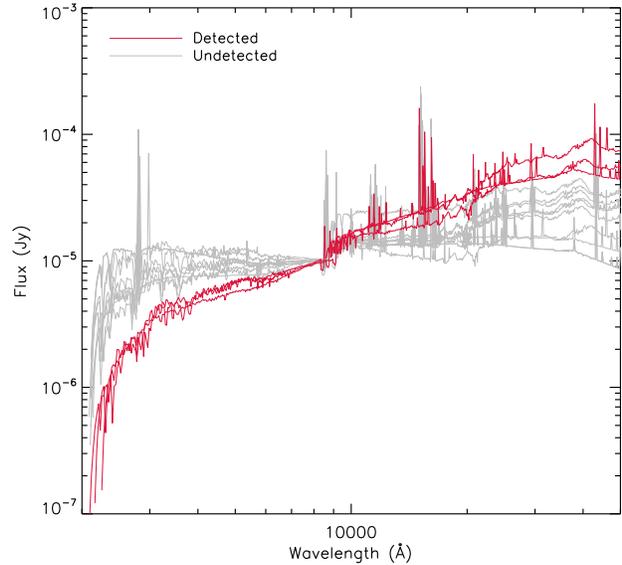}
\end{center}
\vspace*{-0.2cm}\caption[UV to NIR SED fits for detected vs, non-detected sources]{Ultraviolet to near infrared SED fits for detected (red) and non-detected sources (gray) reveal that the three detected sources have the reddest UV SEDs of the WiggleZ kinematic sample. Small variations in the flux level of the SEDs are normalised to a common flux value at $8\times10^3 \AA$. }
\label{fig.redseds}
\end{figure}

The three detected galaxies have the largest \halpha sizes ($r_{1/2} \geq 4$ kpc), the highest expected dust extinctions ($A_\mathrm{V}\geq 0.7$ mag) and are massive ($\log(M_*[\Msun]) \geq 10.7$). They do not correlated to the highest \halpha SFR or specific SFR objects.
The three detections show the largest velocity shear within the sample corresponding to two disc galaxies and one merger. None of the compact ($r_{1/2}\sim 1-2$ kpc) or `dispersion dominated' galaxies were detected. The other non-detected disc galaxies all have sizes of $\leq 3$ kpc and estimated global dust extinctions of $\leq 0.5$ mag.
The implications of the kinematics of the detections vs. non detections is discussed in more detail in Section~\ref{sec5.discussion}.
As expected, and in agreement with the PACS detected LBGs at $z=0.8-1.2$ \citep{2013arXiv1306.1121O} the WiggleZ SPIRE detections are also the reddest objects in the sample as probed by their UV-NIR SEDs shown in Fig.~\ref{fig.redseds}.

\subsection{Dust properties}
\label{subsec5.blackbody}

\begin{figure*}
\begin{center}\hspace*{-0.5cm}\includegraphics[width=0.42\textwidth, angle=90]{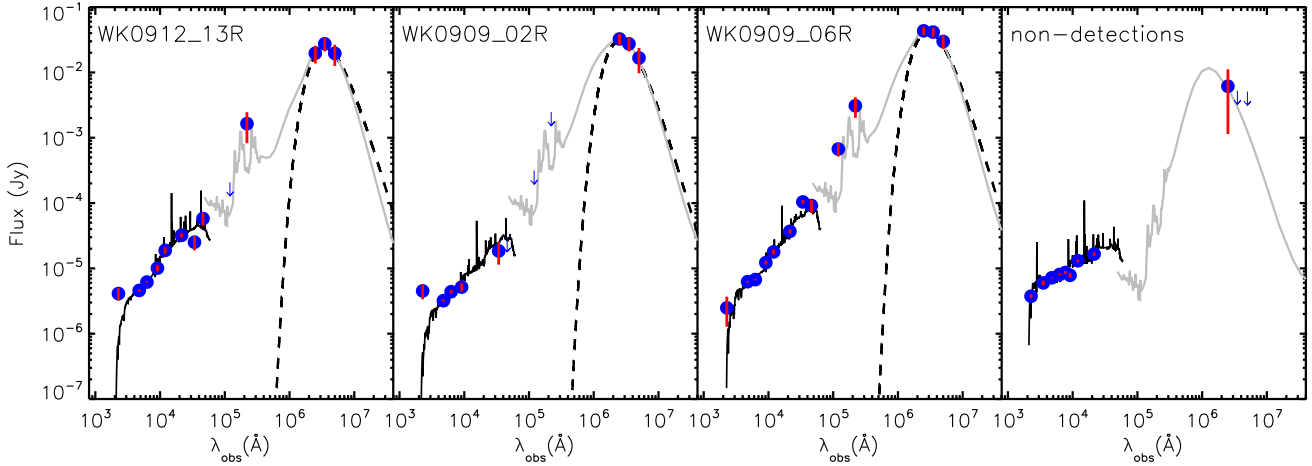}
\end{center}\caption[FIR spectral energy distributions of the three detected WiggleZ sources]{Infrared spectral energy distributions of the three detected WiggleZ sources, WK0912\_13R, WK0909\_02R, and WK0909\_06R, and the average spectral energy distribution of the non-detections. The blue points are the observed UV to FIR fluxes of the galaxies and downward pointing arrows indicated limits. The dashed black line is the best-fit graybody curve to the SPIRE photometry used to derive dust properties, as described in Section~(\ref{subsec5.blackbody}). SED fits to the UV to NIR from \cite{2011MNRAS.417.2601W} and SED fits to the MIR to FIR using templates from \cite{2002ApJ...576..159D} are shown by the solid black and gray curves, respectively.}  
\label{fig.seds}
\end{figure*}

To obtain infrared luminosities and dust temperatures a blackbody emission curve in the optically thin approximation, or graybody, was fit to the measured SPIRE fluxes given in Table~\ref{table.herschel} defined by,
\begin{equation}
S_{\nu} \propto \frac{\nu^{3+\beta}}{exp(h\nu/kT_\mathrm{d})-1},
\label{eq.graybody}
\end{equation}
where $S_{\nu}$ is the observed flux density at frequency $\nu$, $h$ is the Planck constant, $k$ is the Boltzmann constant, \Td~is the dust temperature, and the dust emissivity spectral index is set to $\beta=1.5$. A degeneracy exists between the dust emissivity and dust temperature in the case where only limited frequency measurements are available, as such $\beta $ is held constant \citep{2003MNRAS.338..733B}. The value of $\beta$ is not well constrained, ranging from $1.5-2.0$ for high-redshift galaxies. Our results are relatively insensitive to the choice of $\beta$ with the difference in \Td[K] of $\Delta$\Td~$=4$ and a difference in \LIR~of $\Delta\log$(\LIR[\Lsun]) $=0.04$ when $\beta=2.0$, comparable to the statistical errors. This simple model has been shown to give reliable measurements of dust properties for sparsely sampled SEDs (e.g. \citealt{2002PhR...369..111B,2003MNRAS.338..733B,2006ApJ...650..592K,2006MNRAS.370.1185P}). 

For the WiggleZ detections at $z\sim1.3$, the three SPIRE bands constrain the turnover of the graybody providing a reliable estimate of the dust temperature. The best-fit graybody is shown in Fig.~\ref{fig.seds} for the three detected WiggleZ sources, with values obtained for \Td~= [21.6, 29.8, 27.5] K, and $\log$(\LIR[\Lsun]) = [11.8, 12.1, 12.3] for  WK0912\_13R, WK0909\_02R, and WK0909\_06R respectively.  Far-infrared SEDs were fit to the WISE and Herschel photometry to show the overall shape of IR SED using the \cite{2002ApJ...576..159D} templates, which are shown in Fig.~\ref{fig.seds} with the UV-NIR photometry and SED fits. 

A Monte-Carlo analysis is performed to measure errors on \Td~and \LIR~by varying the SPIRE fluxes within a Normal distribution around the measured value. The derived values for WK0912\_13R and WK0909\_06R are well determined as the three SPIRE points probe the full turn over.  The derived values for WK0909\_02R are much more uncertain, $\sigma$(\Td) $\sim9$ K, as the measured 250 $\mu$m flux less clearly probes the Wein side of the blackbody curve. We note here that the results from staking analysis indicate that the non-detections have warmer dust as the 250 \mum,~350 \mum~and 500 \mum~imply a descending slope as expected on the Wein side of the blackbody curve.

The stacking analysis of Section~\ref{subsec5.stacking} yields a detection for the 250 \mum~band and upper limits for the two long wavelength bands suggesting that on average the SPIRE bands probe the Rayleigh-Jeans side of the graybody curve for the non-detections. This assumption allows for a determination of lower limits on the dust temperature and infrared luminosity of \Td~$>27$ K and $\log$(\LIR[\Lsun])~$>11.3$ respectively. Figure 4 shows the average SED for the non-detections including the stacking results. One possible FIR SED model is shown from \cite{2002ApJ...576..159D} for comparison to the SEDs of the detected galaxies. 

Comparison samples in the WiggleZ detections redshift range ($1.2<z<1.6$) with FIR measurements include star-forming galaxies (LBGs: \citealt{2011ApJ...734L..12B}; `main-sequence' galaxies: \citealt{magnelli:2012aa}), starbursts \citep{casey:2012ab}, and SMGs \citep{2004ApJ...617...64S,2010MNRAS.409L..13C,2012arXiv1202.0761M}, where the value of dust emissivity is constant across these samples ($\beta=1.5$). The starbursts and SMGs have average \Td~$\sim30-50$ K and \LIR~$\gtrsim10^{12.5}$ \Lsun.  More typical star-forming galaxies are considerably more difficult to detect, as reiterated by this work. As a result there are only four individual galaxies at this redshift for comparison, with \Td$\sim 34$ K and \LIR~$\sim 6\times10^{11}$ \Lsun~\citep{2011ApJ...734L..12B,magnelli:2012aa}. A comparison with these samples suggests that derived infrared properties for the WiggleZ sample (from the detections and implied from the non-detections), are reflective of `average' or main-sequence systems, in agreement with the conclusions drawn from their \halpha properties \citep{2011MNRAS.417.2601W}. 
The differences of \LIR~and \Td~between galaxy populations at $z\sim1.4$ are consistent with studies at lower redshifts and in high-redshift survey fields, where a more statistical study of the \Td$-$\LIR~plane is possible (e.g. \citealt{2003ApJ...588..186C,2010MNRAS.409...75H,Roseboom:2011fk,2012arXiv1202.0761M, 2013MNRAS.431.2317S}). However, these distinctions (particularly at $z>1$), can be biased by the detection limits of current and past infrared facilities (e.g. \citealt{2012arXiv1202.0761M,casey:2012ab}).

Dust masses are estimated from the modified blackbody such that, 

\begin{equation}
M_\mathrm{dust} = \frac{S_{\nu} D_L^2}{(1+z)\kappa_\mathrm{rest}B_{\nu}(\lambda_\mathrm{rest}, T_\mathrm{dust})},
\label{eq.dustmass}
\end{equation}
where the dust mass absorption coefficient $\kappa_\mathrm{rest}$ is given by
\begin{equation}
\kappa_\mathrm{rest} = \kappa_\mathrm{obs}(\frac{\lambda_\mathrm{obs}}{\lambda_\mathrm{rest}})^\beta,
\label{eq.dustmass2}
\end{equation}
$B_{\nu}$ is the Planck function at frequency $\nu$, and $D_L$ represents the luminosity distance. Following \cite{li:2001aa}, $\kappa_\mathrm{obs}$ is set to 0.517 m$^2$ kg$^{-1}$ for observed wavelength 250 $\mu$m. 
The estimated dust masses are log($M_\mathrm{dust}$[\Msun]) = [9.3, 9.4, 9.0] for WK0912\_13R, WK0909\_02R, and WK0909\_06R respectively. Statistical errors are measured using the Monte Carlo techniques described above and are given in Table~\ref{table.herschel}. These values are also uncertain due to the assumptions of both $\beta$ and $\kappa_\mathrm{obs}$. The choice of $\beta=2$, rather than $\beta=1.5$ used here, results in higher dust masses such that $\Delta \log(M_\mathrm{dust})=0.1$, within the statistical errors. Furthermore, the blackbody fitting method used to estimate the dust mass has been shown to give masses a factor of 2 lower than an alternative method using the models from \cite{2007ApJ...657..810D} \citep{magdis:2012aa}. 

Recent works have derived gas masses by assuming that gas mass is proportional to dust mass at a given metallicity (e.g. \citealt{2011ApJ...737...12L,magdis:2012aa,magnelli:2012aa}). While these methods are uncertain (e.g. due to the reliability of metallicity estimates and assuming simple dust model used here) they are applied to the WiggleZ detections to derive approximate gas masses and thus gas fractions. Using the relation from \cite{2011ApJ...737...12L},
\begin{equation}
\log\delta_\mathrm{GDR} = (9.4\pm1.1)-(0.85\pm0.13)[12 + \log(\mathrm{O/H})],
\end{equation}
and the metallicities derived from the \NII/\halpha from \cite{2011MNRAS.417.2601W}, gas-to-dust ratios, $\delta_\mathrm{GDR}\equiv M_\mathrm{gas}/M_\mathrm{dust}$, are estimated in the range 135-160. Gas fractions, $f_\mathrm{gas}$, are estimated as $\sim$0.86, 0.94, and 0.57 for WK0912\_13R, WK0909\_02R, and WK0909\_06R respectively where $f_\mathrm{gas}=M_\mathrm{gas}/(M_\mathrm{gas}+M_*)$. At $z\sim1.5$ the estimated values for WiggleZ galaxies are in broad agreement with gas fractions for main sequence galaxies at the same redshift as estimated from CO emission \citep{2010ApJ...713..686D, 2010Natur.463..781T}.

\subsection{Correcting for dust extinction}
\label{sec5.dust}
Observed luminosities and star formation rates derived from the \halpha and infrared observations are compared in order to estimate the attenuation from intervening dust and to bring the near-infrared and far-infrared global SFR indicators into agreement. Infrared luminosities are commonly used to derive more accurate estimates of the true SFR of a galaxy by assuming that all \halpha light attenuated by dust is re-emitted in the infrared \citep{1990ApJ...350L..25D,2001ApJ...555..613I}. This assumption is dependent on the nature (and uniformity) of the dust and the stellar population responsible for heating the dust. Nonetheless, assuming that infrared radiation for SFGs results from young stars and the stellar radiation field \citep{2000ApJ...533..682C}, the infrared luminosities for the WiggleZ galaxies are used to estimate reddening and dust attenuation. 

Total SFRs as derived from the infrared have traditionally been estimated using the theoretical prescription of \cite{1998ApJ...498..541K} for \LIR[8-1000 $\mu$m].  However, this prescription was designed for starbursts (continuous bursts of age $10-100$ Myr) and although it has been shown to be accurate for a range of galaxy populations \citep{2002AJ....124.3135K}, it can systematically over-estimate total SFRs \citep{2009ApJ...703.1672K,nordon:2010aa}. Results from recent infrared facilities have led to tighter conversions from \LIR~to SFR. Unfortunately, most new SFR prescriptions require information from single wavebands particular to each infrared facility (e.g. 8 \mum~or 24 \mum) which are not available for the WiggleZ sample (e.g. \citealt{2005ApJ...633..871C,2007ApJ...667L.141R,2009ApJ...692..556R}). However, these have been brought together for a generalised total IR/\halpha SFR indicator by combining IRAS and \textit{Spitzer} data for star-forming galaxies at $z=0$ from the SINGS survey \citep{2009ApJ...703.1672K}.

This prescription is used to estimate the dust attenuation and total SFRs using \LIR~and \Lha~for the WiggleZ kinematic sample.  It is noted that (1) both prescriptions are calibrated on local star-forming galaxies and thus may not be applicable to this $z\sim1.3$ sample and (2) as no continuum is detected in the OSIRIS observations an assumption is made that there is no underlying \halpha stellar absorption.
This method uses an empirical prescription to estimate dust extinction in the form of,
\begin{equation}
A_{\mathrm{H}\alpha} = 2.5\log \left [ 1+ \frac{a_\lambda L_\lambda}{L_{\mathrm{H}\alpha}}  \right ],
\label{eq.kennicut1}
\end{equation}
which can be used to calculate a dust corrected luminosity from a linear combination of \halpha luminosity and luminosities from a variety of infrared indicators, 
\begin{equation}
L_{\mathrm{H}\alpha}^\mathrm{corr} = L_{\mathrm{H}\alpha}^\mathrm{obs} + a_\lambda L_\lambda,
\label{eq.kennicut2}
\end{equation}

\begin{equation}
SFR_{\mathrm{H}\alpha}^\mathrm{corr} [ \Msun~\mathrm{yr}^{-1}] =  7.9\times10^{-42} L_{\mathrm{H}\alpha}^\mathrm{corr} [\mathrm{ergs~s}^{-1}]
\label{eq.kennicut3}
\end{equation}
where $\lambda$ in this case is represented by the total infrared luminosity, TIR[3-1100 \mum], with the corresponding scaling coefficient, found empirically to be $a_\mathrm{TIR}=0.0024\pm0.0006$ and the attenuation in \halpha represented by $A_{\mathrm{H}\alpha}$. The definition of TIR[3-1100 \mum] \citep{2009ApJ...703.1672K} differs slightly from the adopted range in this paper, IR[8-1000 \mum], this slight difference is assumed to have a negligible effect on $a_\mathrm{TIR}$.

\begin{figure}
\begin{center}\hspace*{-0.5cm}\includegraphics[scale=0.5]{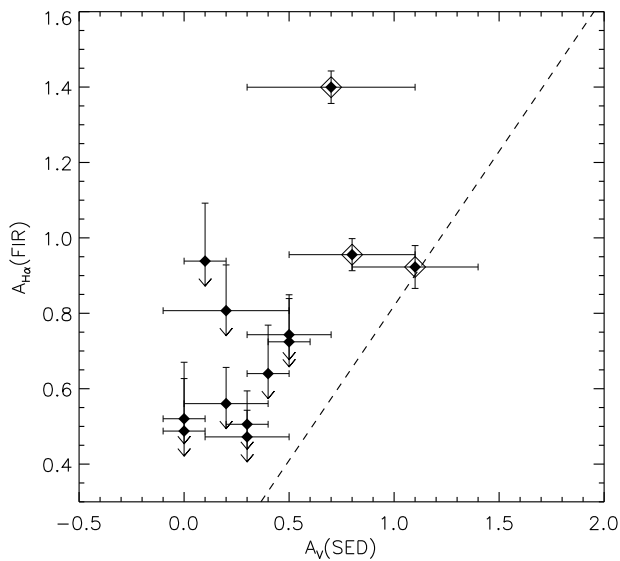}
\end{center}\caption[Comparison of extinctions derived from SED fitting and FIR-\halpha]{Comparison of extinctions  $A_V$, derived from UV-optical-NIR SED fitting, and $A_{\mathrm{H}\alpha}$ derived from \LIR/L$_{\mathrm{H}\alpha}$ using the empirical results of \citet{2009ApJ...703.1672K}. The dashed line represents the relationship between $A_V$ and $A_{\mathrm{H}\alpha}$ using the galactic obscuration curve of \citet{2000ApJ...533..682C}. The detections are represented by outlined diamonds. The upper limits for the non-detections are derived from fitting modified blackbody curves to $3\times\sigma_\mathrm{conf}$ holding both \Td~constant (top of  $A_{\mathrm{H}\alpha}$ error bar) and \LIR~constant (diamond) to the average values of these parameters derived from the detections. } 
\label{fig.dust2}
\end{figure}

As expected the SFR derived directly from the \halpha gives a lower estimate of the total SFR than the far-infrared, with SFR$_\mathrm{IR}$/SFR$_{\mathrm{H}\alpha}\sim4$, using the \cite{1998ARA&A..36..189K,1998ApJ...498..541K} SFR conversions.  The \cite{2009ApJ...703.1672K} prescription brings the two star formation rate indicators ($L_{\mathrm{H}\alpha}^\mathrm{corr}$ and \LIR) in better agreement within a factor of SFR$_\mathrm{IR}$/SFR$^\mathrm{corr}_{\mathrm{H}\alpha}\sim1.5$. The new SFR values, SFR$_{\mathrm{H}\alpha}^\mathrm{corr}$, are given in Table~\ref{table.herschel}. One possible contributor to the discrepancy between the two indicators is the relatively low metallicity of the WiggleZ galaxies, 12+log(O/H)$\sim8.5$, which could bias the ratio of SFRs \citep{2012MNRAS.426.1782R,dominguez:2012:05}.  However it is noted, the galaxy with the most discrepant \LIR~and $L_{\mathrm{H}\alpha}$ is WK0909\_02R. Optical imaging of WK0909\_02R indicates that the OSIRIS \halpha luminosity is underestimated due to an aperture effect as multiple components of the system cover $\sim4$ arc seconds, comparatively larger than the field of view of OSIRIS. In contrast, the SPIRE beam would encompass the whole system, accounting for the difference in SFR indicators. This aperture effect should be negligible for WK0909\_06R, and WK0912\_13R as they appear to be disc galaxies with the bulk of \halpha detected within the OSIRIS field of view. 

The FIR derived $A_{\mathrm{H}\alpha}$ values of the three WiggleZ galaxy detections (A$_\mathrm{V}\sim1-2$ mag) are comparable to other $z>1$ galaxy extinction measurements (e.g. \citealt{1999MNRAS.306..843G,2013arXiv1307.3556I}) and in line with A$_\mathrm{V}\sim1$ mag normally assumed for local SFGs \citep{1992ApJ...388..310K,1996MNRAS.281..847T,2001PASP..113.1449C}, whereas dust in more extreme systems, i.e. SMGs, can suppress the observed star formation rate by a factor of 10 \citep{2004ApJ...617...64S}.

In comparison, the dust correction derived from the UV-NIR SEDs leave the corrected \halpha SFRs over a factor of SFR$_{\mathrm{H}\alpha}^\mathrm{corr, IR}$/SFR$^\mathrm{corr,SED}_{\mathrm{H}\alpha}\sim1.5$ discrepant. This comparison can be seen in Fig.~\ref{fig.dust2} between far-infrared derived $A_{\mathrm{H}\alpha}$ and UV-NIR SED derived $A_V$.
For the WiggleZ galaxies, the attenuation derived by SED fitting are systematically underestimated, as the three detected objects have the highest values of dust attenuation from the UV-NIR fits, $A_V>0.6$ mags.

\section{Kinematic implications of \textit{Herschel} observations}
\label{sec5.discussion}
From \halpha morphologies, the three detected objects are classified as `multiple emission' galaxies with two, WK0909\_06R and WK0912\_13R, being the largest disc galaxies in the sample containing resolved star-forming regions, reminiscent of local \HII regions \citep{2012MNRAS.422.3339W}. No `compact emission' galaxies were detected. 
The results of this analysis could be used to support the conclusion that the multiple emission galaxies are mergers based upon, 1.) mergers are a common mechanism to drive-up \LIR~$-$ e.g. ULIRGs and SMGs 2.) the clumpy structure could be a result of dust variation across the galaxy rather than the true structure of the young stars, and 3.) there is some overlap of the \LIR-\Td~values between the WiggleZ results and SMGs. However, WK0909\_06R and WK0912\_13R have infrared luminosities of $\sim 1\times10^{12}$ L$_{\odot}$, an order of magnitude lower than most typical $z\sim2$ SMGs that are known to be mergers. Some observations also suggest that some `cold' SMGs ($<20$ K) are contaminated by nearby sources, biasing luminosity derivations and underestimating temperatures \citep{2010MNRAS.409...75H}. Yet, at this \LIR~the fractional contribution of spirals is expected to be less (25\%) than of major mergers (40\%). These percentages however are based on a wide redshift range $z=0.1-3.5$ and thus likely integrates over an evolution in luminosity and galaxy type \citep{2010ApJ...721...98K}. Combining the kinematic results and \halpha clump analysis for the disc galaxies it is more likely that these results indicate that the far-infrared emission originates from the same young stellar populations that generates the \halpha emission, namely the star-forming clumps identified in \cite{2012MNRAS.422.3339W} which enhance the stellar radiation field in the multiple emission galaxies, increasing the likelihood of detection with \textit{Herschel} (e.g. \citealt{2000A&A...363....9Z,2002AJ....124.3135K}). The other disc-like galaxies, which do not show large clumps, are undetected.

However, following this assumption a detection is expected for WK1002\_61S, the only other galaxy classified as having multiple regions of emission in this sample, which has an \halpha luminosity $\sim2\times$ the \halpha luminosity of WK0909\_02R and WK0912\_13R. The galaxy WK1002\_61S has $2\sigma_\mathrm{conf}$ detections in the 350~\mum~and 500~\mum~SPIRE bands. It is possible that inclination may have an effect. However, from \halpha morphologies this object looks most edge-on of the four multiple emission galaxies, implying a greater dust extinction (e.g. \citealt{1975gaun.book..123H,2007ApJ...659.1159S}). Deep optical broad-band imaging is needed to get a more complete morphological picture of this object.

None of the compact ($r_\mathrm{eff}\sim 1$ kpc) `single emission' galaxies are detected with \textit{Herschel}.  A possible interpretation is that these galaxies are consistent with undergoing a starburst or AGN mode. If these objects were starbursts resultant from mergers they would be expected to lie above the `main sequence' of star-formation, be more compact than main sequence galaxies, and have warmer dust temperatures (\Td~$\sim$ 40 K; \citealt{2011A&A...533A.119E}). These properties are broadly consistent with what is observed here, however the infrared luminosity of the non-detections is not constrained and thus their offset from the main-sequence is unknown. Although these galaxies do not show properties of hosting an AGN in their optical-NIR spectra \citep{2011MNRAS.417.2601W}, they may contain dust enshrouded AGN. However, the observed wavebands of this study do not probe the mid infrared where the broadband SED of AGN systems peak \citep{2010A&A...518L..33H,2012ApJ...759..139K}. 

Alternatively, If the infrared luminosities of the non-detected compact single emission galaxies are low, $\sim 5\times10^{11}$ \Lsun, with dust temperatures of $\sim30$ K (just below our detection limit) then they may be consistent with pseudo-bulges of young discs possibly forming after the coalescence of clumps in clumpy discs \citep{Elmegreen:2008fk}. Indeed, the single-clump objects show the highest velocity dispersions (Table~\ref{table.global5}) which is qualitatively in agreement with the bulge formation models of \citep{Bournaud:2009xz}. Another possible scenario is that they are compact discs that are unresolved even with adaptive-optics aided $\sim750$ pc resolution \citep{newman:2012aa}.

The kinematic/FIR results presented here are in agreement with what has been derived from morphologies of star-forming galaxies with FIR detections \citep{magnelli:2012aa,2013arXiv1306.1121O}.  The only other sample with published \halpha IFS kinematics and FIR properties at $z>1$ is a submillimeter sample of three galaxies. These galaxies at $z=[1.4, 2.2, 2.3]$, originally detected at 850 $\mu$m with SCUBA, have hot dust temperatures of $\sim35-55$ K and infrared luminoities of $9-60\times10^{12}$ \Lsun~\citep{2013arXiv1302.2145M,2005ApJ...622..772C}. The \halpha properties and kinematics support the hypothesis that the objects are mergers or starbursts showing broad \halpha line widths, high star formation surface densities and no evidence for rotation.

\section{Conclusions}
\label{sec5.conclusions}
In the UV-selected WiggleZ kinematic sample, three of thirteen galaxies are detected at the 3$\sigma$ level with \textit{Herschel}/SPIRE. The dust SEDs of the three detected galaxies are dominated by cold dust (\Td $\sim26$ K) consistent with the ISM heating by the UV and interstellar radiation field. Modified blackbody fits reveal average infrared luminosities and dust masses of \LIR~$\sim 1.2\times10^{12}$ \Lsun~and $M_\mathrm{dust}\sim9.2$ \Msun~respectively, yielding inferred gas fractions of $>50\%$. The FIR results of the sample are in agreement with the interpretation of the NIR \halpha kinematic data, showing the strength of combing multi-wavelength data with integral field spectroscopy for a more complete understanding of the kinematics and formation of individual galaxies.

In this sample, two of the detected sources are likely disc galaxies and one a starburst resulting from a merger. The two disc candidates have the largest spatial extent within the sample and contain multiple regions of emission, suggesting that the cold dust detected is a result of heating of the interstellar radiation field by the clumps. The other detection is more likely to be a starburst resulting from an ongoing merger. This assumption is supported by broadband imaging and long-slit spectroscopy that confirms that there are two close companions to the brightest knot of \halpha emission. The compact single emission galaxies, not detected in the FIR. Stacking analysis reveals that they likely have warmer dust temperatures, \Td~$>27$, on average than the detected objects.  The combination of the kinematic and FIR data suggest that they may be pseudo-bulges in a later phase of clumpy disc formation, however a starburst or AGN mode can not be ruled out.

The data presented here are among the few individual main-sequence star-forming galaxies to be detected at high redshift in the far infrared with \textit{Herschel}-SPIRE \citep{2011A&A...533A.119E,2011ApJ...734L..12B,magnelli:2012aa}. They confirm the stacked results of \cite{2012ApJ...744..154R} that typical UV-selected galaxies have analogous dust properties as local LIRGs and LBGs at $z>1$ \citep{2011ApJ...734L..12B} and have comparable infrared properties to $z\sim1$ samples of star-forming galaxies \citep{dominguez:2012:05,magnelli:2012aa}. The average attenuation of the detected galaxies is $A_{\mathrm{H}\alpha}=1.1$ mag or $A_V=0.9$ mag, assuming a \cite{2000ApJ...533..682C} obscuration curve, leading to dust corrected \halpha star-formation rates of $\sim3\times$ the observed \halpha star formation rates. The non-detections have an attenuations $A_{\mathrm{H}\alpha}<0.76$ mag, when fixing the dust temperature.

Resolved dust properties are a key measurement needed to understand the thick clumpy discs at high redshift and the complexity and impact of mergers. With the end of the \textit{Herschel} mission the next step forward will be the comparison of FIR properties from ALMA with $z>1$ resolved galaxy kinematics. Future kinematic surveys may also find overlap with detections from the deeper Herschel surveys, especially at $z\lesssim1$. The upcoming KMOS$^\mathrm{3D}$ survey in the deep fields is one such example.

\section*{Acknowledgments}
We would like the thank the referee for valuable comments. EW is very grateful to B. Magnelli, D. Lutz, and J. T. Mendel for useful conversations. AMS gratefully acknowledges an STFC Advanced Fellowship through grant ST/H005234/1. 

This paper used data from the \textit{Herschel Space Telescope}. The Herschel spacecraft was designed, built, tested, and launched under a contract to ESA managed by the Herschel/Planck Project team by an industrial consortium under the overall responsibility of the prime contractor Thales Alenia Space (Cannes), and including Astrium (Friedrichshafen) responsible for the payload module and for system testing at spacecraft level, Thales Alenia Space (Turin) responsible for the service module, and Astrium (Toulouse) responsible for the telescope, with in excess of a hundred subcontractors.
\bibliographystyle{apj}


%
\clearpage

\clearpage
\label{lastpage}

\end{document}